\documentclass[10pt]{article}
\usepackage[a4paper, left=1in, right=1in, top=1in, bottom=1in]{geometry}
\usepackage[affil-it]{authblk}
\usepackage{multirow}
\usepackage{graphicx}
\newif\ifsubmissionimages
\submissionimagestrue
\ifsubmissionimages
  
\else
  
\fi
\usepackage{amsmath,amssymb}
\usepackage{xcolor}
\usepackage{arydshln}
\usepackage[ruled,vlined]{algorithm2e}
\usepackage[numbers,sort&compress]{natbib}
\usepackage{xurl}
\usepackage[pdfborder={0 0 0}, colorlinks=true, citecolor=black,  linkcolor=black, urlcolor=blue]{hyperref}
\usepackage[font=small,labelfont=bf]{caption}
\usepackage[section]{placeins}
\usepackage{caption}
\begin{document}
\raggedbottom

\title{\Large\bf When is vaccine prioritization worth optimizing?}

\author[1,4]{Mi Feng}
\author[6]{Zhaohua Lin}
\author[1,2,3]{Changsong Zhou\thanks{cszhou@hkbu.edu.hk}}
\author[1,3,5]{Liang Tian\thanks{liangtian@hkbu.edu.hk}}

\affil[1]{Department of Physics, Hong Kong Baptist University, Kowloon Tong, Hong Kong SAR 999077, China}
\affil[2]{Centre for Nonlinear Studies and Beijing-Hong Kong-Singapore Joint Centre for Nonlinear and Complex Systems (Hong Kong), Hong Kong Baptist University, Kowloon Tong, Hong Kong SAR 999077, China}
\affil[3]{Institute of Computational and Theoretical Studies, Hong Kong Baptist University, Kowloon, Hong Kong SAR 999077, China}
\affil[4]{Institute for Research and Continuing Education, Hong Kong Baptist University, Shenzhen, Guangdong, 518057, China}
\affil[5]{Institute of Systems Medicine and Health Sciences, Hong Kong Baptist University, Kowloon Tong, Hong Kong SAR 999077, China}
\affil[6]{School of Medical Imaging, Fujian Medical University, Fuzhou, Fujian, 350108, China}
\renewcommand\Authands{, }
\date{}
\maketitle

\begin{abstract}
    
    Optimizing vaccine prioritization is often treated as the default policy response when vaccine supply is limited.
    Yet optimized prioritization carries administrative, ethical and communication costs, motivating an upstream question:
    whether differences among vaccine allocations can alter epidemic outcomes enough to make optimization epidemiologically necessary.
    We show that optimization is not always worth pursuing: in some regimes, vaccination markedly reduces epidemic burden, 
    but many feasible allocation rules perform almost equally well, making the necessity of optimization low.
    We quantify this necessity as the range of epidemic outcomes generated by different allocations under fixed supply
    and show that it is governed by competition between vaccinating high-contact groups to slow transmission and vaccinating groups that benefit most directly:
    necessity is low when these protection routes are balanced and high when one dominates.
    Increasing transmission intensity changes this balance and drives a transition in the optimal allocation from transmission-focused prioritization toward direct protection.
    Different prevention objectives exhibit distinct transition thresholds, 
    creating regimes in which optimizing one objective substantially compromises another,
    thereby revealing when the choice of prevention target matters most.
    This framework reframes vaccine prioritization as a prior decision problem, 
    identifying when optimization is warranted, when simpler rules suffice, and when prevention goals conflict.

\end{abstract}

\section*{Introduction}

When vaccine supply is scarce, allocating doses among population groups becomes unavoidable~\cite{soltesz2020effect,pollard2021guide}, 
and optimizing vaccine prioritization can appear to be the natural policy response. 
In this study, we show that optimization is not always epidemiologically necessary, a distinction motivated by two linked considerations.
First, implementing finely targeted prioritization schemes often requires detailed population data, complex logistics, and coordinated communication,
and may raise concerns about fairness, acceptability, and public trust~\cite{chen2022strategic,ismail2020framework,akbarpour2024economic,georgiadis2021optimal}.
Second, in some epidemiological regimes, these costs may yield little additional epidemiological benefit: 
under fixed vaccine supply, many feasible allocation rules produce nearly indistinguishable outcomes relative to one another, 
even though vaccination itself substantially reduces epidemic burden. 
In such regimes, the incremental gain from selecting the optimized allocation is small, making the epidemiological necessity of optimization low. 
This raises a consequential upstream but underexplored question: under what epidemiological conditions can different vaccine allocations produce sufficiently different epidemic outcomes to warrant optimization?

\begin{figure}
	\centering
	\includegraphics[width=\textwidth]{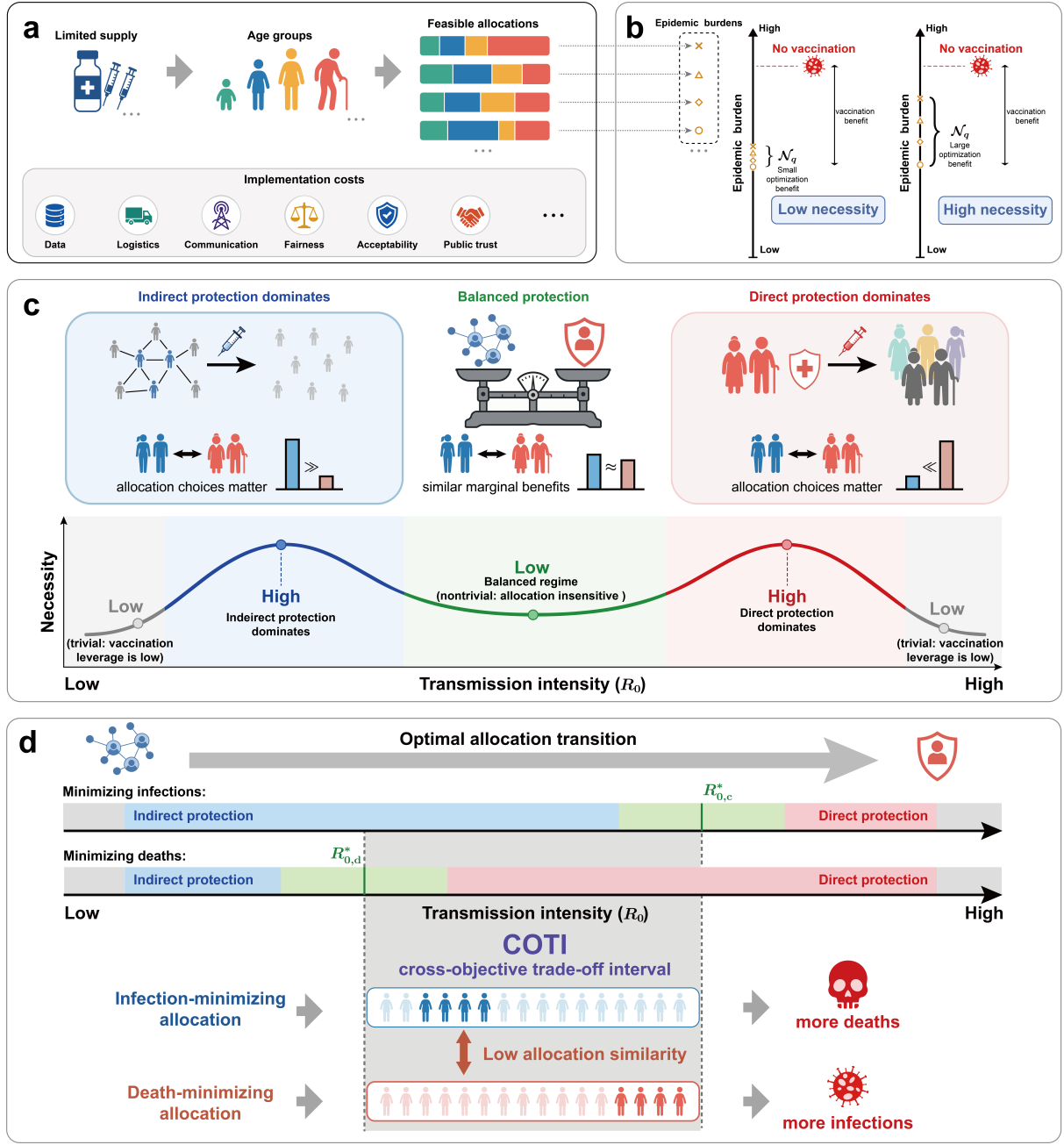}
	\caption{
		\textbf{Epidemiological necessity as a prior decision framework for vaccine prioritization.}
		\textbf{a} Limited vaccine supply makes allocation across age groups unavoidable and generates multiple feasible allocations under the same supply. 
		Implementing finely targeted prioritization also entails costs related to population data, logistics, communication, fairness, acceptability, and public trust.
		\textbf{b} Under fixed supply, feasible allocations map to epidemic burdens. For prevention objective $q$, epidemiological necessity $\mathcal{N}_q$ is the range between the minimum and maximum attainable burdens across feasible allocations. 
		A narrow range indicates low necessity, whereas a wide range indicates high necessity. 
		The difference between no vaccination and the optimized allocation represents the benefit of vaccination itself, whereas $\mathcal{N}_q$ represents the added benefit available from optimizing how the fixed supply is distributed.
		\textbf{c} Epidemiological necessity is governed by the trade-off between indirect and direct protection. 
		Indirect protection vaccinates high-contact groups to reduce onward transmission, whereas direct protection vaccinates groups for whom vaccination most directly reduces the target burden. 
		Necessity is low in the balanced regime, where the marginal benefits of the two protection routes are similar and allocation choices have similar effects on epidemic burden. 
		Necessity is high when one protection route dominates, because allocation choices that move doses away from the dominant route can markedly increase epidemic burden. 
		Low-necessity labels at the two transmission extremes mark boundary references where vaccination has little leverage; the main low-necessity mechanism highlighted here is the interior balanced-protection regime, where vaccination remains effective but optimization adds little.
		\textbf{d} As $R_0$ increases, the optimal allocation transitions from transmission-focused prioritization toward direct protection. 
		Different prevention objectives can undergo this transition at distinct thresholds, creating a cross-objective trade-off interval (COTI) in which objective-specific allocation patterns have low alignment and optimizing one prevention target can compromise another.}
	\label{fig:illustration}
\end{figure}

Most vaccine-prioritization studies begin after a key policy decision has already been made: that allocation should be optimized. 
They have generated sophisticated answers to the downstream question of how doses should be allocated once targeted prioritization is pursued. 
Model-informed studies have compared age-based, risk-based, serostatus-based, occupation-based, and dynamically updated strategies for reducing infections or deaths under limited supply~\cite{medlock2009optimizing,matrajt2010optimizing,bubar2021model,matrajt2021vaccine,buckner2021dynamic}. 
These studies provide rich guidance for designing allocation policies under limited supply, but they rarely ask whether the epidemiological gain from optimization is large enough to justify the social, administrative, and economic costs of implementing an optimized strategy.
Motivated by such broader policy concerns, a related line of work has broadened the allocation problem by incorporating social utility, equity, feasibility, ethical principles, and economic externalities~\cite{chen2022strategic,schmidt2021equitable,ismail2020framework,akbarpour2024economic,emanuel2020fair,world2020sage}. 
For example, Chen et al. showed that behavior- and demography-aware allocation can improve both social utility and equity by prioritizing disadvantaged communities~\cite{chen2022strategic}. 
Applying such frameworks, however, requires aggregating epidemiological outcomes with administrative effort, ethical acceptability, public communication, trust, and feasibility. 
These considerations are difficult to quantify, highly context-dependent, and often contested, and the value weights assigned to them vary across settings and decision-makers, 
limiting generalizability and practical interpretability~\cite{emanuel2020fair,emanuel2023shared,ismail2020framework,world2020sage}. 
A criterion that does not require assigning weights to social and implementation costs is therefore needed to assess the epidemiological consequence of allocation choices before optimization is pursued, 
so that one can determine whether optimization is epidemiologically necessary before broader policy trade-offs are introduced.
Fig.~\ref{fig:illustration}a situates this criterion in the policy setting: limited supply creates feasible age-group allocations, while finely targeted prioritization entails implementation costs that motivate a pre-optimization diagnostic.

We formalize this question through epidemiological necessity $\mathcal{N}_q$ for a given prevention objective $q$, 
defined as the range of final epidemic burdens attainable when the same vaccine supply is allocated across age groups in different feasible ways. 
For objectives such as reducing cumulative infections or deaths, $\mathcal{N}_q$ provides a global measure of allocation sensitivity: 
a wide range means that epidemic burden can change substantially with how doses are distributed, so optimization can produce a large additional reduction in final burden,
whereas a narrow range indicates that different feasible allocations produce similar outcomes, so the incremental gain from selecting the optimized allocation is small. 
In this sense, epidemiological necessity separates the benefit of vaccination itself from the added benefit of optimizing how a fixed supply is distributed. 
This measurement logic is illustrated in Fig.~\ref{fig:illustration}b: limited supply defines a set of feasible allocations across age groups, and epidemiological necessity is represented visually by the burden range generated by those allocations under fixed supply.

When vaccination can still markedly reduce epidemic burden,
epidemiological necessity arises from the trade-off between two protection routes: indirect and direct protection~\cite{halloran1997study,fine2011herd}. 
Indirect protection is achieved by vaccinating high-contact groups, thereby reducing onward transmission and indirectly lowering infections across the population. 
Direct protection instead prioritizes groups for whom vaccination most directly reduces the target burden, 
such as groups likely to benefit before infection or groups facing higher fatality risk. 
Necessity is low when these two protection routes are balanced: 
in such a regime, the marginal benefits of vaccination are similar across age groups, 
so shifting doses from one group to another has little effect on epidemic burden, leaving many allocation rules close to the optimum. 
Necessity is high when one protection route dominates, with its marginal benefit clearly exceeding that of the other; 
reallocating doses away from the dominant route can produce a large increase in epidemic burden.
Fig.~\ref{fig:illustration}c illustrates this mechanism, demonstrating how the balance between protection routes determines epidemiological necessity.

Transmission intensity can turn vaccine prioritization into a conflict between prevention goals.
As the basic reproduction number $R_0$ increases, the optimal allocation undergoes a transition from transmission-focused prioritization toward direct protection. 
Prevention objectives, such as reducing cumulative infections or deaths, can have different thresholds at which this transition occurs. 
For instance, the allocation optimized for reducing deaths can move toward direct protection before the allocation optimized for reducing cumulative infections.
This separation creates an intermediate transmission regime in which the allocations favored by different objectives diverge sharply. 
Within this regime, using the allocation optimized for one prevention target can materially compromise another.
For example, the infection-minimizing allocation can produce more deaths than the death-minimizing allocation, 
while the death-minimizing allocation can yield more infections than the infection-minimizing allocation.
We measure this loss by asking how well each objective-specific allocation performs under the alternative objective, 
revealing when prevention objectives cannot be treated as interchangeable.
Fig.~\ref{fig:illustration}d schematizes this mechanism, showing how separated objective-specific transitions create a cross-objective trade-off interval.

Taken together, our study turns vaccine prioritization from a search for a single optimum into a prior decision problem: 
before asking which allocation is optimal, it asks whether prioritization is worth optimizing. 
Epidemiological necessity identifies when different allocation rules lead to sufficiently different epidemic burdens to support complex prioritization, and when simpler rules may suffice. 
The same framework also quantifies when an allocation optimized for one prevention goal, such as reducing cumulative infections, imposes a large loss on another goal, such as reducing deaths. 
This study provides a practical basis for avoiding unnecessary optimization when allocation rules perform similarly, 
prioritizing careful implementation when allocation choices materially change epidemic outcomes, 
and recognizing trade-offs among competing prevention goals when objective-specific allocations diverge.

\section*{Results}
\setcounter{subsection}{0}
\renewcommand{\thesubsection}{\arabic{subsection}}

\subsection{Quantifying epidemiological necessity under fixed supply}

We quantify epidemiological necessity by asking how much the final epidemic burden can change when the same vaccine supply is distributed in different ways. 
To evaluate the final burden for each feasible allocation, we use an age-structured non-Markovian extension of the Susceptible--Infected--Removed (SIR) model 
that incorporates vaccination through vaccinated ($V$) and protected ($P$) states, yielding a Susceptible--Infected--Removed--Vaccinated--Protected (SIRVP) model~\cite{feng2019equivalence,feng2025dynamic,sherborne2018mean,starnini2017equivalence,feng2023validity,lin2020non,lin2024higher}.
This model maps each age-specific vaccine allocation $\boldsymbol{\vartheta}$ to a final epidemic burden $\tilde{\chi}_q(\boldsymbol{\vartheta})$ for prevention objective $q$; 
see Supplementary Sections~1--2 for the notation and for the full compartmental structure, event-time distributions, vaccination update, and final-state prediction.
Here, final epidemic burden refers to the burden at the final state, when the outbreak has ended and no new infections occur~\cite{kermack1927contribution,andreasen2011final,breda2012formulation}. 
We consider two objectives in the main analysis, $q\in\{\mathrm{c},\mathrm{d}\}$, where $\mathrm{c}$ denotes cumulative infections and $\mathrm{d}$ denotes deaths, as described in Methods; 
the cumulative-infection burden at the final state is also known as the final size. 
This allocation-to-burden mapping provides the basis for defining epidemiological necessity under fixed supply.

Let $\Theta$ denote the feasible set of age-specific vaccine allocations under a fixed-supply constraint (see Methods for the detailed definition). 
For each $\boldsymbol{\vartheta}\in\Theta$, the total vaccine supply is fixed, whereas its distribution across age groups can vary. 
For a given prevention objective $q$, we define two objective-specific bounding allocations over the same feasible set:
the burden-minimizing allocation is the feasible allocation that attains the lowest final burden, and, following common usage, we refer to it as the optimal allocation;
the burden-maximizing allocation is the feasible allocation that attains the highest final burden. 
These allocations are obtained as:
\begin{align}
	\boldsymbol{\vartheta}^{\mathrm{opt}}_{q}
	&=
	\arg\min_{\boldsymbol{\vartheta}\in\Theta}
	\tilde{\chi}_{q}(\boldsymbol{\vartheta}), \label{eq:opt_allocation}\\
	\boldsymbol{\vartheta}^{\mathrm{max}}_{q}
	&=
	\arg\max_{\boldsymbol{\vartheta}\in\Theta}
	\tilde{\chi}_{q}(\boldsymbol{\vartheta}).
	\label{eq:max_allocation}
\end{align}
These constrained optimization problems are solved numerically as described in Methods.
When $q=\mathrm{c}$, we refer to these allocations as the infection-minimizing and infection-maximizing allocations, with ``infection'' denoting final cumulative infections. 
When $q=\mathrm{d}$, we refer to them as the death-minimizing and death-maximizing allocations. 
These bounding allocations are objective-specific: the infection-minimizing and infection-maximizing allocations generally differ from their death-minimizing and death-maximizing counterparts.

\begin{figure}[!tb]
	\centering
	\includegraphics[width=\textwidth]{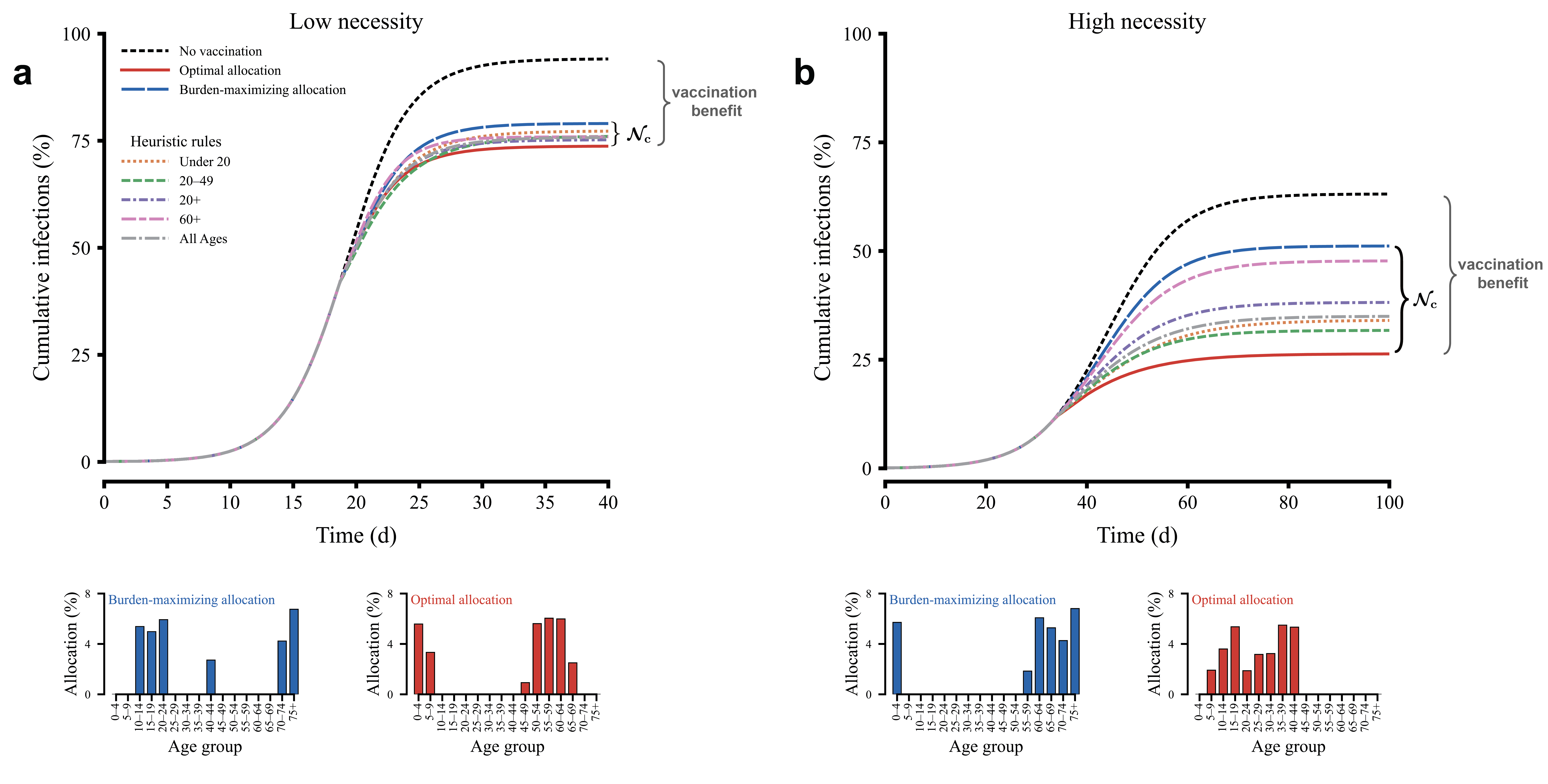}
	\caption{
		\textbf{Representative low- and high-necessity regimes for minimizing cumulative infections.}
		The examples use cumulative infections as the prevention objective, so epidemiological necessity is denoted by $\mathcal{N}_{\mathrm{c}}$. 
		Here, the two bounding allocations are the infection-minimizing allocation, referred to as the optimal allocation, and the infection-maximizing allocation, referred to as the burden-maximizing allocation; ``infection'' denotes final cumulative infections.
		Unless otherwise stated in the main-text analyses, vaccine efficacy is $\eta=95\%$, vaccine supply is $B=30\%$ of the total population, and vaccination is implemented when cumulative infections have reached $10\%$ of the increase from the initial state to the no-vaccination final state.
		\textbf{a} In the low-necessity regime, vaccination substantially reduces cumulative infections relative to no vaccination, but the optimal and burden-maximizing allocations reach similar final burdens. 
		The resulting upper--lower necessity $\mathcal{N}_{\mathrm{c}}$ is small, and common heuristic rules perform similarly to the optimal allocation.
		\textbf{b} In the high-necessity regime, feasible and heuristic allocations produce widely separated final burdens. 
		The optimal allocation attains the lower final-burden bound, whereas the burden-maximizing allocation attains the upper bound, yielding a large $\mathcal{N}_{\mathrm{c}}$.
		The ``All ages'' heuristic denotes the population-proportional allocation, following the terminology of Bubar et al.~\cite{bubar2021model}.
		The lower bar plots beneath each trajectory show the age-specific vaccine distributions for the burden-maximizing and optimal allocations, making explicit how the same total supply is redistributed across age groups in the two bounding allocations. 
		The burden-maximizing allocation is used as the upper-bound reference for defining the attainable burden range, rather than as a policy recommendation.
		Gray brackets indicate the reduction from no vaccination to the optimal allocation, representing the benefit of vaccination itself; black brackets indicate $\mathcal{N}_{\mathrm{c}}$, the burden range attributable to allocation choices under fixed supply.
	}
	\label{fig:illustration_example}
\end{figure} 

We define epidemiological necessity for objective $q$ as the upper--lower burden range across feasible allocations under the same vaccine supply:
\begin{align}
	\mathcal{N}_{q}
	=
	\tilde{\chi}_{q}\!\left(\boldsymbol{\vartheta}^{\mathrm{max}}_{q}\right)
	-
	\tilde{\chi}_{q}\!\left(\boldsymbol{\vartheta}^{\mathrm{opt}}_{q}\right).
	\label{eq:necessity}
\end{align}
Because all allocations in $\Theta$ use the same vaccine supply, differences in $\tilde{\chi}_{q}(\boldsymbol{\vartheta})$ arise solely from how that supply is distributed across age groups. 
Thus, $\mathcal{N}_q$ measures the full allocation-induced burden range within the fixed-supply feasible set, without requiring a particular unoptimized reference allocation, thereby providing a benchmark-free global measure of allocation sensitivity.
When this range is wide, allocation choices can materially change epidemic burden and optimization is epidemiologically consequential; 
when it is narrow, feasible allocations yield similar burdens and optimization adds little beyond simpler feasible rules.
The main qualitative features of the necessity landscapes are robust to alternative operational definitions of necessity:
the same qualitative landscape is obtained when necessity is defined relative to a population-proportional benchmark (Supplementary Section~3). 

Fig.~\ref{fig:illustration_example} provides the intuition behind this definition using cumulative infections as the prevention objective, so that $q=\mathrm{c}$ and epidemiological necessity is denoted by $\mathcal{N}_{\mathrm{c}}$. 
In the low-necessity example (Fig.~\ref{fig:illustration_example}a), vaccination markedly reduces cumulative infections relative to no vaccination, but the optimal and burden-maximizing allocations reach similar final burdens. 
Because the common heuristic rules shown here also satisfy the fixed-supply constraint, their final burdens lie within the same narrow attainable range and remain close to the optimal allocation, making the additional gain from optimization small. 
In the high-necessity example (Fig.~\ref{fig:illustration_example}b), feasible allocations generate a broad range of final burdens. 
The optimal allocation produces a much lower final burden than the burden-maximizing allocation, and common heuristic rules can lie far from the optimum within the attainable range. 
A large $\mathcal{N}_{\mathrm{c}}$ therefore indicates that allocation choices are epidemiologically consequential.
These examples distinguish the overall benefit of vaccination from the epidemiological necessity of optimizing how a fixed vaccine supply is distributed: a large vaccination benefit does not imply high optimization necessity. 
Across transmission regimes, feasible allocations of the same supply can generate either narrow or wide final-outcome ranges, corresponding to low or high epidemiological necessity. 
Supplementary Sections~3 and 4 further reinforce this distinction: the main necessity landscape is robust to benchmark-based definitions, whereas vaccination-benefit landscapes depend strongly on which vaccinated allocation is chosen as the reference. 
Together, these results show that the qualitative structure of optimization necessity reflects allocation sensitivity within the fixed-supply vaccinated range and remains distinct from the reference-specific patterns of vaccination benefit.

In the main analysis, necessity is evaluated under a one-time vaccination campaign, which provides a clean and theoretically tractable setting for diagnosing allocation sensitivity; the rationale for this choice is discussed further in the Discussion. 
A complementary time-course rollout analysis tests this modeling choice and shows that the conclusions developed below remain robust when allocations are optimized and evaluated under time-course vaccine delivery (Supplementary Section~5).

\subsection{Balanced protection lowers necessity for minimizing cumulative infections}

\begin{figure}[!tb]
	% \centering
	\includegraphics[width=\textwidth]{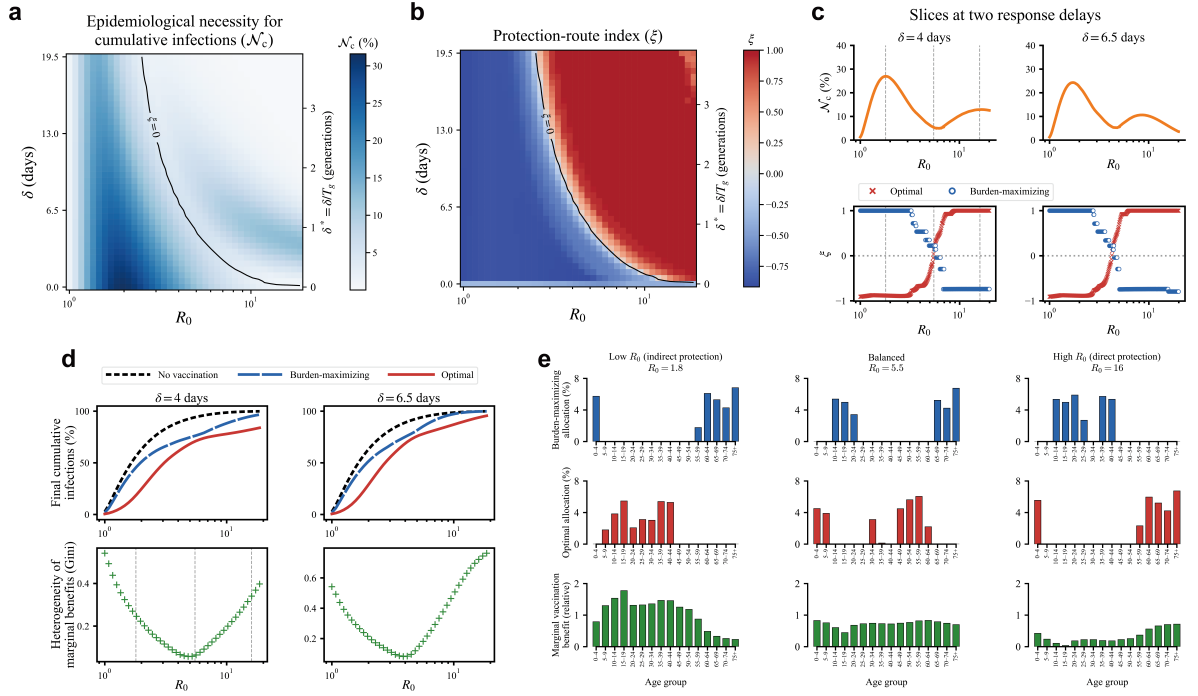}
	\caption{
		\textbf{Balanced protection lowers epidemiological necessity for minimizing cumulative infections.}
		\textbf{a} Heat map of epidemiological necessity $\mathcal{N}_{\mathrm{c}}$ over transmission intensity $R_0$ and vaccine-response delay $\delta$. 
		For fixed $\delta$, necessity is low in boundary regimes where vaccination has limited leverage, high when one protection route dominates, and low near the balanced-protection regime,
		with the black contour reproduced from (\textbf{b}).
		\textbf{b} Heat map of the protection-orientation index $\xi^{\mathrm{opt}}_{\mathrm{c}}$ of the infection-minimizing allocation over the same $(R_0,\delta)$ plane. 
		The contour $\xi^{\mathrm{opt}}_{\mathrm{c}}=0$ marks the transition between indirect- and direct-protection routes; the low-necessity valley in panel (\textbf{a}) lies near this transition region. 
		Right ticks in panels (\textbf{a}) and (\textbf{b}) indicate the rescaled delay $\delta^*=\delta/T_g$ in generations, with $T_g=5.05$ days under the COVID-19 calibration.
		\textbf{c} Fixed-delay slices at $\delta=4$ and $6.5$ days.
		The top row plots $\mathcal{N}_{\mathrm{c}}$ as a function of $R_0$, and the bottom row reports the protection-orientation indices $\xi^{\mathrm{opt}}_{\mathrm{c}}$ and $\xi^{\mathrm{max}}_{\mathrm{c}}$ for the infection-minimizing and infection-maximizing allocations. 
		Vertical dashed lines mark the representative transmission regimes illustrated in panel (\textbf{e}). 
		\textbf{d} Final cumulative infections and marginal-benefit heterogeneity along the same two delay slices. 
		The top row compares the final burdens under no vaccination, the infection-minimizing allocation, and the burden-maximizing allocation. 
		The bottom row reports the Gini coefficient of age-specific marginal vaccination benefits; larger values indicate stronger heterogeneity in the marginal benefit of vaccinating different age groups. 
		\textbf{e} Representative age-specific profiles at indirect-oriented, balanced, and direct-oriented regimes of optimal allocation. 
		Rows show, from top to bottom, the burden-maximizing allocation, the optimal allocation, and the age-specific marginal vaccination benefit. 
		As transmission increases, the optimal allocation shifts from high-contact groups, through a mixed balanced profile, toward groups with stronger direct benefits, while the burden-maximizing allocation follows the opposite orientation. 
		Marginal benefits are concentrated in high-contact groups in the indirect-oriented regime, become more evenly distributed near the balanced regime, and shift toward groups with stronger direct benefits in the direct-oriented regime. 
	}
	\label{fig:necs_cinf}
\end{figure}

We first ask when optimizing vaccine allocation is epidemiologically necessary for reducing cumulative infections. 
To answer this question, we vary the basic reproduction number $R_0$, which controls transmission intensity, and the vaccine-response delay $\delta$, defined as the interval between vaccine administration and the onset of vaccine-induced protection. 
During this delay, vaccinated-but-not-yet-protected individuals remain susceptible to infection, so $\delta$ controls how quickly vaccine-induced protection develops after vaccination. 
The numerical value of $\delta$ should be interpreted relative to the epidemic clock of the calibrated model, rather than as a universal biological time-to-protection across outbreak time scales (see Methods).
Across this parameter space, the resulting necessity landscape is strongly regime-dependent (Fig.~\ref{fig:necs_cinf}a; Fig.~\ref{fig:necs_cinf}c, top row).
At the two transmission boundaries, $\mathcal{N}_{\mathrm{c}}$ is low: when $R_0$ is close to $1$, outbreaks have limited growth potential, whereas when $R_0$ is extremely large, vaccination has limited ability to alter the final burden regardless of how doses are allocated. 
Between these extremes, where vaccination can alter epidemic outcomes, $\mathcal{N}_{\mathrm{c}}$ exhibits two high-necessity regions separated by a low-necessity valley. 
This intermediate valley is the regime of primary interest because vaccination remains effective while different allocations yield similar final burdens.

The intermediate valley is explained by a balance between indirect and direct protection. 
We summarize whether an allocation is oriented toward indirect or direct protection using the protection-orientation index $\xi$, 
defined in Methods as a comparison between contact-driven and benefit-driven reference directions 
(written, when needed, as $\xi^{\mathrm{opt}}_{\mathrm{c}}$ for the infection-minimizing allocation, with superscripts denoting allocation type and subscripts denoting prevention objective).
Under this sign convention, negative values indicate indirect protection, positive values indicate direct protection, and values near zero mark a balance between the two routes. 
Indirect protection prioritizes high-contact groups to reduce onward transmission, whereas direct protection prioritizes groups that are more likely to acquire vaccine-induced protection before infection.
For cumulative infections, lower-contact groups are more likely to remain uninfected until vaccine-induced protection develops; for deaths, this direct-protection effect is reinforced by the higher fatality risk of older groups. 
As a result, in both objectives, direct protection corresponds to prioritizing older, lower-contact groups.
As $R_0$ increases, the optimal allocation shifts from indirect protection toward direct protection 
(Fig.~\ref{fig:necs_cinf}b; Fig.~\ref{fig:necs_cinf}c, bottom row; representative allocation profiles in Fig.~\ref{fig:necs_cinf}e, second row).
The $\xi^{\mathrm{opt}}_{\mathrm{c}}=0$ contour lies close to the interior low-necessity valley 
(Fig.~\ref{fig:necs_cinf}a,b), indicating that $\mathcal{N}_{\mathrm{c}}$ decreases when the optimal allocation passes through a balanced-protection regime.

This protection-orientation transition explains why allocation choices matter in some regimes but not in others. 
At lower values of $R_0$, the epidemic is relatively easy to control, and vaccinating high-contact groups can effectively suppress transmission, allowing indirect protection to confer broader benefits.
At higher values of $R_0$, transmission is harder to contain, and indirect effects weaken; thus, prioritizing groups with strong direct benefits is more effective in reducing the final epidemic burden.
In both regimes, the optimal strategy emphasizes one dominant effect, 
whereas the burden-maximizing allocation emphasizes the opposite effect;
this separation creates a large final-burden range and yields high epidemiological necessity (excluding the two extreme boundary regimes; Fig.~\ref{fig:necs_cinf}c, bottom row; Fig.~\ref{fig:necs_cinf}e, top two rows).
When the optimal strategy balances the two effects, the burden-maximizing allocation also remains balanced, so the two bounding allocations produce similar final burdens and $\mathcal{N}_{\mathrm{c}}$ decreases 
(Fig.~\ref{fig:necs_cinf}c, bottom row; Fig.~\ref{fig:necs_cinf}e, top two rows). 

Two forms of low necessity should be distinguished (Fig.~\ref{fig:necs_cinf}d, top row). 
At extremely low or extremely high $R_0$, low $\mathcal{N}_{\mathrm{c}}$ mainly reflects limited vaccination leverage: 
the epidemic either has little capacity to grow or is too intense for vaccination to substantially change the final size (Fig.~\ref{fig:necs_cinf}d). 
This boundary low-necessity regime is therefore trivial because optimization adds little when vaccination itself has a limited effect. 
By contrast, the intermediate low-necessity valley is the more informative case. 
There, vaccination still reduces cumulative infections, but feasible allocations produce similar final burdens; 
optimization is less necessary not because vaccination is ineffective, but because the final burden is less sensitive to how the fixed supply is distributed.

Age-specific marginal vaccination benefits provide a local explanation for this reduced allocation sensitivity 
(Fig.~\ref{fig:necs_cinf}d, bottom row; Fig.~\ref{fig:necs_cinf}e, bottom row).
We define the marginal benefit of an age group as the reduction in epidemic burden produced by assigning an infinitesimal dose to that group, 
and summarize its heterogeneity by the Gini coefficient, both defined in Methods with the Jacobian derivation given in Supplementary Section~2.
When $R_0$ is low, marginal benefits concentrate in high-contact groups; when $R_0$ is high, they shift toward groups with stronger direct benefits. 
In both cases, the benefit distribution is heterogeneous, so reallocating doses can produce large changes in final burden. 
The Gini coefficient decreases near the balanced regime, and the marginal-benefit profiles become flatter across age groups.
This flattening means that shifting doses from one group to another has a smaller effect on cumulative infections, explaining why balanced protection lowers $\mathcal{N}_{\mathrm{c}}$.

This cumulative-infection pattern is robust under time-course vaccine rollout and under perturbations of population and contact profiles (Supplementary Sections~5--6).
It also retains the same qualitative organization under age-dependent vaccine efficacy and under sensitivity analyses of vaccine efficacy, vaccination timing and vaccine supply (Supplementary Sections~7--8).

\subsection{Age-specific objective weights reshape necessity for death minimization}
\begin{figure}[!tb]
	\centering
	\includegraphics[width=\textwidth]{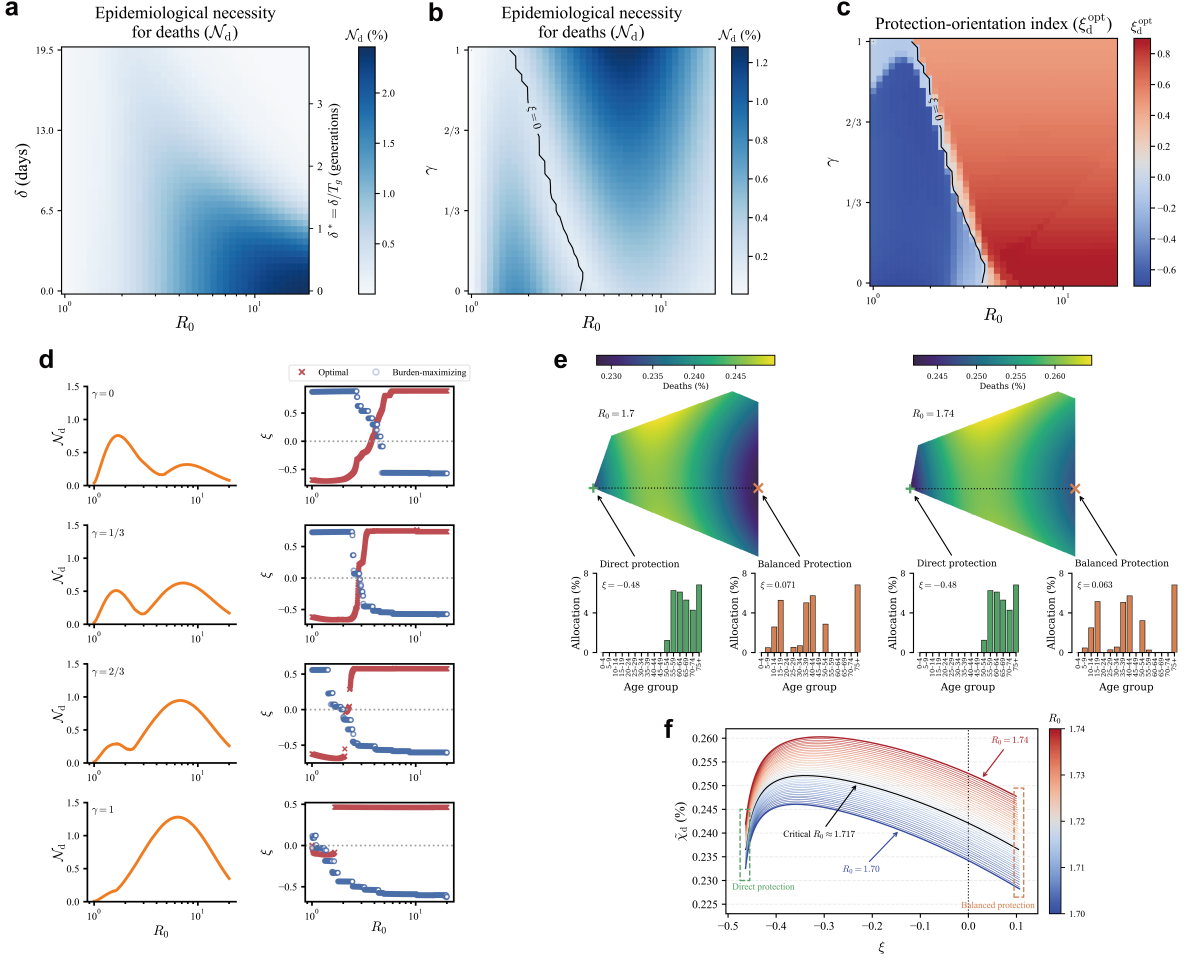}
	\caption{
	\textbf{Age-specific objective weights reshape necessity for death minimization.}
	\textbf{a} Heat map of epidemiological necessity for minimizing deaths, $\mathcal{N}_{\mathrm{d}}$, over transmission intensity $R_0$ and vaccine-response delay $\delta$. 
	Under the baseline age-specific IFR profile, the dependence on $R_0$ is primarily unimodal for fixed $\delta$: necessity is concentrated in an interior transmission regime rather than forming the two-peak structure observed for cumulative infections. 
	Right ticks indicate the rescaled response delay $\delta^*=\delta/T_g$ in generations, with $T_g=5.05$ days under the COVID-19 calibration. 
	\textbf{b} Epidemiological necessity $\mathcal{N}_{\mathrm{d}}$ over transmission intensity $R_0$ and IFR-heterogeneity coefficient $\gamma$, with $\delta$ fixed at $7$ days. 
	The coefficient $\gamma$ interpolates between a homogeneous IFR profile ($\gamma=0$) and the highly heterogeneous baseline age-specific IFR profile ($\gamma=1$). 
	The black contour is the $\xi^{\mathrm{opt}}_{\mathrm{d}}=0$ contour reproduced from panel (\textbf{c}). 
	As IFR heterogeneity increases, the low-$R_0$ necessity peak weakens and the high-$R_0$ direct-protection peak becomes dominant. 
	\textbf{c} Heat map of the protection-orientation index $\xi^{\mathrm{opt}}_{\mathrm{d}}$ of the death-minimizing allocation over the same $(R_0,\gamma)$ plane. 
	The contour $\xi^{\mathrm{opt}}_{\mathrm{d}}=0$ marks the route-balance transition. 
	\textbf{d} Fixed-$\gamma$ slices of $\mathcal{N}_{\mathrm{d}}$ and the protection-orientation indices as functions of $R_0$ for $\gamma\in\{0,1/3,2/3,1\}$. 
	The left column plots $\mathcal{N}_{\mathrm{d}}$, and the right column reports $\xi^{\mathrm{opt}}_{\mathrm{d}}$ and $\xi^{\mathrm{max}}_{\mathrm{d}}$ for the death-minimizing and death-maximizing allocations. 
	Increasing IFR heterogeneity suppresses the indirect-oriented peak, strengthens the direct-oriented peak, and changes the optimal-allocation transition from smooth to abrupt. 
	\textbf{e} Local death-burden landscapes near the abrupt transition for $\gamma=1$ and $\delta=7$ days. 
	Each surface is a two-dimensional slice of the allocation space, with color indicating final deaths. 
	The green plus marks a direct-protection local minimum, and the orange cross marks a balanced-protection local minimum; the bar plots below each surface show the corresponding allocation profiles. 
	As $R_0$ increases from $1.70$ to $1.74$, the global optimum switches from the balanced minimum to the direct-protection minimum. 
	\textbf{f} Final deaths along the one-dimensional path connecting the two local minima in panel (\textbf{e}), with curves colored by $R_0$. 
	The crossing of the two minima at $R_0\approx1.717$ produces a discontinuous switch in the death-minimizing allocation. 
	Together, these panels show that IFR heterogeneity reshapes death-minimizing necessity by strengthening direct protection and, at high heterogeneity, creating competing balanced and direct-protection optima.}
	\label{fig:necs_death}
\end{figure}

We then turn from cumulative infections to deaths, where the essential distinction is that deaths define an age-weighted burden objective. 
Cumulative infections weight infections uniformly across age groups; deaths weight them by the age-specific infection fatality rate (IFR). 
Unless otherwise stated, deaths are weighted by a highly heterogeneous baseline age-specific IFR profile, with larger weights assigned to older age groups, derived from reported age-stratified COVID-19 fatality patterns in Refs.~\cite{Actualiz44:online,omori2020age} (Supplementary Section~2).
Thus, the death objective is not only a death-specific target but also a concrete example of how age-specific objective weights reshape allocation sensitivity. 
For fixed vaccine-response delay $\delta$, the death-minimizing necessity $\mathcal{N}_{\mathrm{d}}$ is primarily unimodal as a function of $R_0$ (Fig.~\ref{fig:necs_death}a; Supplementary Section~9). 
This pattern contrasts with the two-peak structure observed for cumulative infections and suggests that risk heterogeneity reshapes allocation sensitivity rather than simply rescaling the necessity profile.

To isolate the role of fatality-risk heterogeneity, we define an IFR-heterogeneity coefficient $\gamma\in[0,1]$ that interpolates between a homogeneous IFR profile and the baseline age-specific IFR profile, as described in Methods and Supplementary Section~2; 
$\gamma=0$ gives homogeneous IFR weights and $\gamma=1$ gives the highly heterogeneous baseline profile.
When $\gamma$ is small, the nearly homogeneous IFR weights make death burden close to a rescaled cumulative-infection burden. 
As a result, the death-minimizing allocation is close to the infection-minimizing allocation, and the necessity profile retains a two-peak structure 
(Fig.~\ref{fig:necs_death}b; Fig.~\ref{fig:necs_death}d, left column).
As $\gamma$ increases, high-IFR age groups receive greater weight in the objective, strengthening the value of direct protection. 
Consequently, the low-$R_0$ indirect-oriented peak weakens, the high-$R_0$ direct-oriented peak grows, and the death-minimizing necessity profile becomes primarily unimodal.

Reshaping also occurs through the transition between protection routes, captured by the protection-orientation index $\xi^{\mathrm{opt}}_{\mathrm{d}}$. 
For weak or moderate IFR heterogeneity, the death-minimizing allocation moves smoothly from indirect protection toward direct protection as $R_0$ increases (Fig.~\ref{fig:necs_death}c; Fig.~\ref{fig:necs_death}d, right column). 
In this regime, the balanced-protection contour $\xi^{\mathrm{opt}}_{\mathrm{d}}=0$ remains aligned with a low-necessity valley, consistent with the mechanism observed for cumulative infections. 
For strong IFR heterogeneity, however, the transition changes qualitatively: the optimal allocation no longer passes smoothly through route balance, but instead switches abruptly from a balanced profile to a direct-protection profile. 

The local geometry of the death objective over allocation space explains this abrupt transition (Fig.~\ref{fig:necs_death}e). 
Near the transition at high IFR heterogeneity, the death-burden landscape contains two competing local minima: one near balanced protection and one near direct protection. 
At $R_0=1.70$, the balanced minimum has the lower death burden and is therefore globally optimal. 
At $R_0=1.74$, the direct-protection minimum becomes lower and takes over as the global optimum. 
Tracking the death burden along the path connecting the two minima shows that their depths cross at $R_0\approx1.717$ (Fig.~\ref{fig:necs_death}f). 
At this crossing, the stability ordering of the two minima reverses, and the global optimum switches discontinuously from balanced protection to direct protection. 

This result clarifies why death-minimizing necessity behaves differently from cumulative-infection necessity. 
For cumulative infections, the objective weights age groups uniformly, so indirect and direct protection compete mainly through transmission intensity and response delay. 
For deaths, IFR heterogeneity adds a second source of structure: directly protecting high-fatality groups becomes increasingly valuable as heterogeneity grows. 
Strong IFR heterogeneity therefore increases the benefits of vaccinating high-risk older groups and shifts the optimal allocation from indirect protection toward direct protection at low $R_0$, 
making balanced protection ($\xi \approx 0$) a local minimum; meanwhile, it lowers the death burden under direct protection, creating a second local minimum there. 
Because this direct-protection pull dominates over a broad range of $R_0$, strong IFR heterogeneity suppresses the indirect-oriented peak and produces the primarily unimodal necessity profile. 

More generally, the death objective illustrates how age-specific objective weights reshape epidemiological necessity. 
Death minimization weights infections by IFR, but other prevention targets may also assign different weights to age groups, such as hospitalization, severe disease, or life-years lost. 
The key determinants are not only the heterogeneity of these weights, but also how high-weight groups align with age-specific transmission roles. 
In the baseline IFR setting, high weights are concentrated in older, lower-contact groups, strengthening direct protection and amplifying the high-$R_0$ necessity peak. 
A counterfactual setting in which high-contact groups carry larger fatality weights reverses this pattern, strengthening the low-$R_0$ indirect-oriented peak instead (Supplementary Section~7). 
Death minimization is therefore a concrete instance of a broader principle: objective weights reshape the protection-orientation landscape, 
and the resulting necessity landscape depends jointly on the heterogeneity of these weights and on the alignment between high-weight groups and high-transmission groups.

This death-objective organization persists under time-course rollout and population- and contact-profile perturbations (Supplementary Sections~5--6).
It also retains its qualitative structure under age-dependent vaccine efficacy and under sensitivity checks for vaccine efficacy, vaccination timing and vaccine supply (Supplementary Sections~7--8).

\subsection{Separated objective-specific transitions create cross-objective trade-offs}

We further examine when the choice of prevention objective becomes epidemiologically consequential. 
The previous sections considered cumulative infections and deaths separately, identifying the allocation that minimizes each objective. 
However, an allocation that is optimal for one objective need not be close to optimal for another. 
An infection-minimizing allocation may cause avoidable deaths, whereas a death-minimizing allocation may reduce deaths at the cost of excess infections.
The key question is therefore not only which allocation is optimal for a given objective, but also how much is lost when that allocation is evaluated under the alternative objective.

\begin{figure}[!tb]
	\centering
	\includegraphics[width=\textwidth]{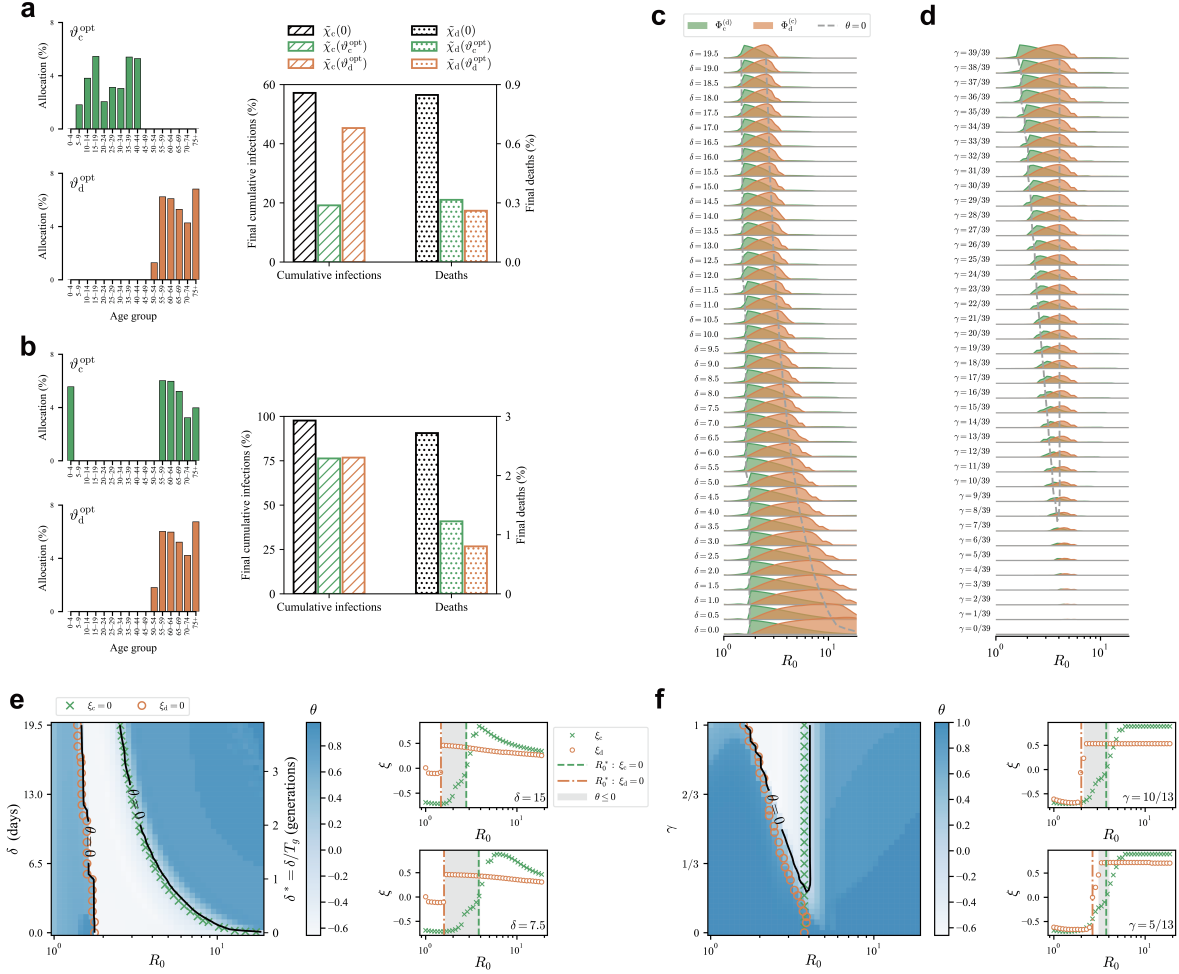}
	\caption{
		\textbf{Objective-specific allocation transitions create cross-objective trade-offs.}
		\textbf{a--b} Representative regimes showing that cross-objective penalties are directional and regime-dependent. 
		In the lower-$R_0$ regime (\textbf{a}), $\boldsymbol{\vartheta}_{\mathrm{c}}^{\mathrm{opt}}$ achieves markedly stronger infection reduction while keeping deaths close to the death-minimizing level; using $\boldsymbol{\vartheta}_{\mathrm{d}}^{\mathrm{opt}}$ therefore sacrifices considerable infection control for only a small additional reduction in deaths. 
		In the higher-$R_0$ regime (\textbf{b}), the asymmetry reverses: using $\boldsymbol{\vartheta}_{\mathrm{c}}^{\mathrm{opt}}$ incurs a larger loss in death reduction, while $\boldsymbol{\vartheta}_{\mathrm{d}}^{\mathrm{opt}}$ sacrifices relatively little infection control.
		\textbf{c--d} Cross-objective penalties as functions of $R_0$ under varying vaccine-response delays $\delta$ (\textbf{c}) and IFR-heterogeneity coefficients $\gamma$ (\textbf{d}). 
		$\Phi_{\mathrm{c}}^{(\mathrm{d})}$ is concentrated toward the lower-$R_0$ side of the COTI; 
		$\Phi_{\mathrm{d}}^{(\mathrm{c})}$ lies toward the higher-$R_0$ side. 
		Increasing $\delta$ shortens the COTI, while increasing $\gamma$ widens it.
		Gray dashed contours mark where the Pearson allocation correlation between $\boldsymbol{\vartheta}_{\mathrm{c}}^{\mathrm{opt}}$ and $\boldsymbol{\vartheta}_{\mathrm{d}}^{\mathrm{opt}}$ is zero.
		\textbf{e--f} Pearson allocation correlation $\theta$ between $\boldsymbol{\vartheta}_{\mathrm{c}}^{\mathrm{opt}}$ and $\boldsymbol{\vartheta}_{\mathrm{d}}^{\mathrm{opt}}$ over the $(R_0,\delta)$ plane (\textbf{e}) and the $(R_0,\gamma)$ plane (\textbf{f}). 
		Right ticks in panel (\textbf{e}) indicate the rescaled delay $\delta^*=\delta/T_g$ in generations, with $T_g=5.05$ days under the COVID-19 calibration.
		Black contours mark the zero-correlation boundary, $\theta=0$, where the two objective-specific allocation profiles cease to be positively aligned.
		Green crosses and orange circles indicate the protection-orientation transition points $\xi_{\mathrm{c}}=0$ and $\xi_{\mathrm{d}}=0$, respectively.
		The side panels show fixed-$\delta$ and fixed-$\gamma$ slices of $\xi_{\mathrm{c}}$ and $\xi_{\mathrm{d}}$ as functions of $R_0$; vertical lines mark the corresponding transition points, and gray shading marks the region where $\theta\le0$.
		The transition points align with the zero-correlation contour, and the $\theta=0$ contours organize the shape of the COTI in (\textbf{c}--\textbf{d}), linking separated objective-specific transitions to the emergence of the COTI.
	}
	\label{fig:cross_obj}
\end{figure}

Two representative regimes reveal opposite cross-objective asymmetries (Fig.~\ref{fig:cross_obj}a--b). 
In a lower-$R_0$ regime, the infection-minimizing allocation $\boldsymbol{\vartheta}_{\mathrm{c}}^{\mathrm{opt}}$ 
produces a much lower cumulative-infection burden than the death-minimizing allocation $\boldsymbol{\vartheta}_{\mathrm{d}}^{\mathrm{opt}}$, 
while still keeping deaths close to the death-minimizing level. 
By contrast, using $\boldsymbol{\vartheta}_{\mathrm{d}}^{\mathrm{opt}}$ provides only a small additional reduction in deaths but sacrifices a large amount of infection reduction (Fig.~\ref{fig:cross_obj}a). 
Thus, the two objective-specific allocations are not equally costly substitutes: 
in this regime, optimizing deaths imposes a large infection-control loss, while optimizing cumulative infections causes only a small loss in death reduction.
At a higher value of $R_0$, the pattern reverses: the infection-minimizing allocation can incur a large loss in death reduction, 
while the death-minimizing allocation sacrifices relatively little infection control (Fig.~\ref{fig:cross_obj}b). 

These two examples motivate a directional measure of cross-objective penalty:
rather than asking only which allocation minimizes its own objective, we ask how much burden increases when an allocation optimized for one objective is evaluated under the other objective.
For two prevention objectives $q$ and $q'$, we define the cross-objective penalty $\Phi_{q}^{(q')}$ as the normalized loss on objective $q$ incurred by using the allocation optimized for objective $q'$:
\begin{align}
    \Phi_{q}^{(q')}=
    \frac{\tilde{\chi}_{q}\!\left(\boldsymbol{\vartheta}_{q'}^{\mathrm{opt}}\right)-\tilde{\chi}_{q}\!\left(\boldsymbol{\vartheta}_{q}^{\mathrm{opt}}\right)}
    {\tilde{\chi}_{q}\!\left(\boldsymbol{0}\right)-\tilde{\chi}_{q}\!\left(\boldsymbol{\vartheta}_{q}^{\mathrm{opt}}\right)},
    \qquad q,q'\in\{\mathrm{c},\mathrm{d}\},\quad q\ne q'.
    \label{eq:cross_objective_penalty}
\end{align}
Here, $\boldsymbol{0}$ denotes the no-vaccination allocation, so $\tilde{\chi}_{q}(\boldsymbol{0})$ is the final burden for objective $q$ in the absence of vaccination.
Thus, $\Phi_{\mathrm{c}}^{(\mathrm{d})}$ measures the proportional loss in cumulative-infection control caused by using the death-minimizing allocation, 
while $\Phi_{\mathrm{d}}^{(\mathrm{c})}$ gives the corresponding loss in death reduction when the infection-minimizing allocation is implemented.
The denominator therefore normalizes the loss by the vaccination benefit available for the evaluated objective, 
so the penalty measures the fraction of that benefit sacrificed by optimizing the alternative objective.

Together, the two penalties define the cross-objective trade-off interval (COTI): the range of $R_0$ in which both $\Phi_{\mathrm{c}}^{(\mathrm{d})}$ and $\Phi_{\mathrm{d}}^{(\mathrm{c})}$ are positive. 
Within this interval, neither objective-specific allocation is interchangeable with the other: optimizing deaths increases cumulative infections, and optimizing cumulative infections increases deaths. 
The penalties have an asymmetric structure across the interval:
the infection penalty $\Phi_{\mathrm{c}}^{(\mathrm{d})}$ is concentrated toward the lower-$R_0$ side, 
and the death penalty $\Phi_{\mathrm{d}}^{(\mathrm{c})}$ lies toward the higher-$R_0$ side (Fig.~\ref{fig:cross_obj}c,d). 
This asymmetry identifies where the choice between infection control and death reduction has the largest consequence.

The COTI emerges from separated objective-specific transitions. 
For both cumulative infections and deaths, indirect protection corresponds to vaccinating high-contact groups, whereas direct protection corresponds here to vaccinating older, lower-contact groups. 
As $R_0$ increases, both objective-specific optima therefore shift toward direct protection, but the death-minimizing allocation shifts earlier because the death objective assigns greater weight to high-fatality groups.
This separation creates an intermediate $R_0$ range in which $\boldsymbol{\vartheta}_{\mathrm{d}}^{\mathrm{opt}}$ has already moved toward direct protection while $\boldsymbol{\vartheta}_{\mathrm{c}}^{\mathrm{opt}}$ remains more indirect-oriented. 
The two optimal allocations then favor different protection directions and become weakly correlated or anticorrelated, as quantified by the Pearson allocation correlation $\theta$ defined in Methods. 
This loss of allocation correlation makes optimization for cumulative-infection control diverge from optimization for death reduction, giving rise to the COTI (Fig.~\ref{fig:cross_obj}e--f). 

This transition-mismatch view also explains how the COTI changes with vaccine-response delay and objective-weight heterogeneity. 
As the vaccine-response delay $\delta$ increases, the infection-minimizing allocation shifts earlier toward direct protection because high-contact groups are more likely to be infected before vaccine-induced protection develops. 
This shift brings $\boldsymbol{\vartheta}_{\mathrm{c}}^{\mathrm{opt}}$ closer to $\boldsymbol{\vartheta}_{\mathrm{d}}^{\mathrm{opt}}$, reducing the transition mismatch and shrinking the COTI. 
By contrast, increasing the IFR-heterogeneity coefficient $\gamma$ strengthens the age-specific weighting in the death objective. 
When $\gamma=0$, deaths are proportional to cumulative infections under homogeneous IFR weights, so the two objectives coincide and the COTI collapses. 
As $\gamma$ increases, the death-minimizing allocation diverges from the infection-minimizing allocation, widening the COTI (Fig.~\ref{fig:cross_obj}c--f).

The COTI identifies the transmission regimes in which prevention objectives should be treated as explicit trade-offs rather than interchangeable targets. 
Outside this interval, optimizing one objective may perform nearly as well under the other, so the practical cost of choosing either target is limited. 
Inside the COTI, objective choice has material epidemiological consequences: a death-focused allocation can sacrifice infection control on the lower-$R_0$ side, while an infection-focused allocation can sacrifice death reduction on the higher-$R_0$ side. 
The COTI therefore marks the regimes in which policymakers should explicitly weigh competing prevention targets, or adopt a composite objective, rather than assuming that one optimized allocation will serve both goals. 

The COTI and directional penalty structure persist under time-course rollout and population- and contact-profile perturbations (Supplementary Sections~5--6), 
and remain qualitatively consistent when fatality weights are made contact-correlated or when vaccine efficacy varies across age groups (Supplementary Section~7).

\subsection{Country-level regime mapping}

\begin{figure}[!b]
	\centering
	\includegraphics[width=\textwidth]{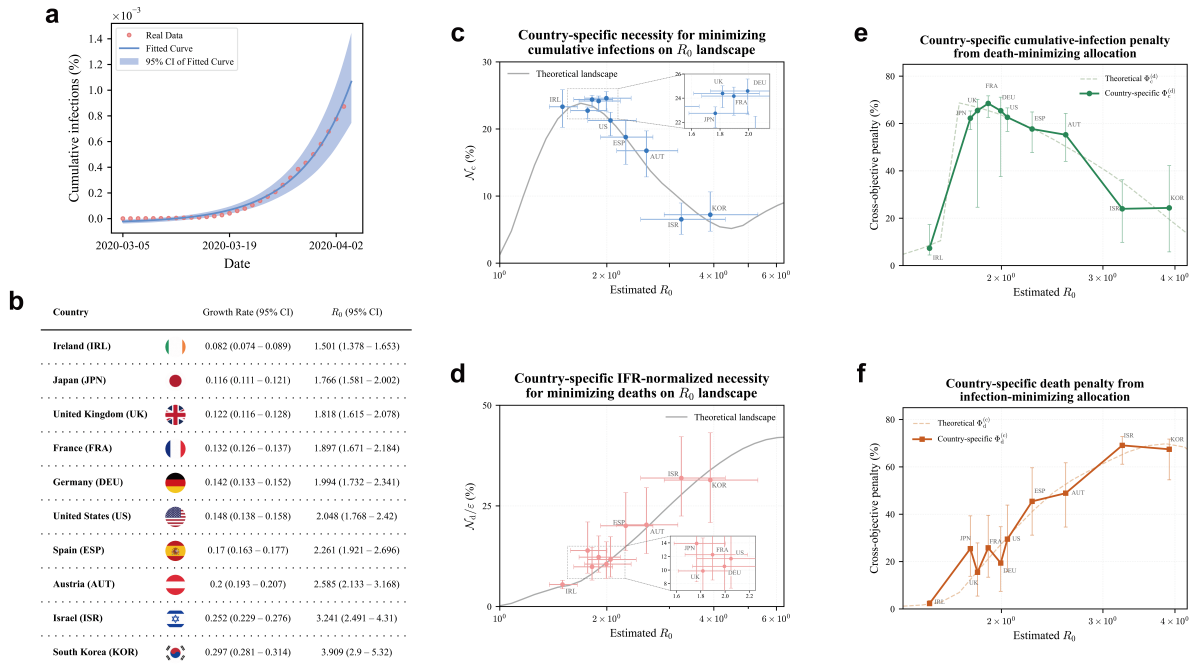}
	\caption{\textbf{Country-level regime mapping during the early stage of the COVID-19 pandemic.}
		\textbf{a} Early cumulative-infection data are fitted with an exponential growth curve to estimate the early growth rate $g$.
		\textbf{b} Estimated $g$ and inferred $R_0$ values, with 95\% uncertainty intervals, for ten countries ordered by increasing median $R_0$.
		\textbf{c} Country-specific cumulative-infection necessity, $\mathcal{N}_{\mathrm{c}}$, projected onto the baseline theoretical $R_0$ profile at $\delta=7$ days. 
		The gray curve shows the baseline theoretical landscape, and country points show uncertainty-propagated estimates from country-specific simulations.
		The estimates mainly occupy the descending branch between the lower-$R_0$ high-necessity regime and the intermediate low-necessity region.
		\textbf{d} Country-specific $\mathrm{IFR}$-normalized death-minimizing necessity, $\mathcal{N}_{\mathrm{d}}/\varepsilon$, projected onto the baseline theoretical $R_0$ profile at $\delta=7$ days, where $\varepsilon$ is the population-weighted overall infection-fatality ratio ($\mathrm{IFR}$) computed from the baseline age-specific IFR profile. 
		The normalization removes the leading population-level fatality-risk scale from the death-minimizing necessity estimates.
		\textbf{e} Cumulative-infection penalty incurred by using the death-minimizing allocation, $\Phi_{\mathrm{c}}^{(\mathrm{d})}$, as a function of inferred $R_0$.
		\textbf{f} Death penalty incurred by using the cumulative-infection-minimizing allocation, $\Phi_{\mathrm{d}}^{(\mathrm{c})}$, as a function of inferred $R_0$.
		In panels \textbf{c--f}, points and error bars summarize medians and uncertainty intervals across the country-specific simulations using $\delta=7$ days. 
		We also repeated the country-specific calculations under a shorter delay, $\delta=3.5$ days, and a longer delay, $\delta=14$ days (Supplementary Section~10). 
		Changing $\delta$ reshapes the corresponding theoretical $R_0$ landscapes and changes which part of each landscape is sampled by the empirical countries, but the country-specific necessity and penalty estimates remain close to the matched theoretical profiles.}
	\label{fig:res_country}
\end{figure}

For each country, we construct a country-specific allocation-to-burden map by replacing the baseline inputs with three country-specific empirical inputs. 
These inputs are the population age profile, the projected age-stratified contact matrix, and the inferred transmission intensity, summarized by $R_0$. 
Population age profiles are taken from Ref.~\cite{WorldPop64:online}, contact matrices from Ref.~\cite{prem2017projecting}, 
and $R_0$ is inferred from country-specific early epidemic growth using COVID-19 generation-time estimates from Refs.~\cite{dong2020interactive,ferretti2020quantifying} (Figs.~\ref{fig:res_country}a--b). 
All other model and vaccination parameters are kept at the baseline values used in the theoretical landscapes, with details of the country-specific parameterization and sensitivity analyses for key fixed parameters provided in Supplementary Section~10.

Across these country-specific simulations, the estimates remain close to the corresponding theoretical $R_0$ landscapes (Figs.~\ref{fig:res_country}c--f). 
This indicates that, even after country-specific population age profiles and contact matrices are included, variation among countries is still largely explained by their inferred transmission intensity. 
This pattern is consistent with the population- and contact-profile robustness analysis in Supplementary Section~6.

For cumulative infections, the ten countries mainly occupy the descending branch between the lower-$R_0$ high-necessity regime and the intermediate low-necessity valley (Fig.~\ref{fig:res_country}c). 
Countries at the lower end of the inferred $R_0$ range, such as Ireland, Japan and the United Kingdom, lie closer to the higher-necessity side of this branch, whereas countries with larger inferred $R_0$, such as Austria, Israel and South Korea, lie closer to the low-necessity region. 

For deaths, raw necessity is amplified in countries with older population profiles because the IFR weights are concentrated in older age groups, so similar allocation-induced infection differences translate into larger death-burden differences.
We therefore compare countries using $\mathcal{N}_{\mathrm{d}}/\varepsilon$, where $\varepsilon=\boldsymbol{\rho}^{\top}\boldsymbol{\epsilon}$ is the population-weighted overall IFR (Fig.~\ref{fig:res_country}d). 
After this normalization, the country estimates follow the baseline death-objective landscape more closely, supporting the interpretation that $\mathcal{N}_{\mathrm{d}}/\varepsilon$ captures the transmission-regime component of death-minimizing necessity rather than the leading fatality-scale difference across populations.

The cross-objective penalties place the same countries within the theoretical trade-off structure (Figs.~\ref{fig:res_country}e--f). 
Most countries fall within the COTI because their inferred $R_0$ values are close to the region where infection-minimizing and death-minimizing allocations diverge. 
Consistent with the theoretical penalty landscape, $\Phi_{\mathrm{c}}^{(\mathrm{d})}$ is larger on the lower-$R_0$ side of the COTI, whereas $\Phi_{\mathrm{d}}^{(\mathrm{c})}$ is larger toward the higher-$R_0$ side. 
Ireland has a relatively low $\Phi_{\mathrm{d}}^{(\mathrm{c})}$, while Israel and South Korea have relatively low $\Phi_{\mathrm{c}}^{(\mathrm{d})}$.

Overall, the country-level analysis links empirical country inputs to the theoretical allocation regimes. 
After replacing the baseline population profile, contact matrix and transmission intensity with country-specific values, each country maps to a corresponding region of the necessity and penalty landscapes. 
Because necessity is measured as a fraction of the total population rather than as an absolute number of people, and because countries differ in population size and policy context, these estimates should not be interpreted as direct rankings of country-level policy urgency. 
Instead, they show where different country settings lie within the theoretical allocation-sensitivity landscape under the same standardized counterfactual vaccination scenario.

\section*{Discussion}
Our framework provides a practical diagnostic for determining whether optimized vaccine prioritization is epidemiologically worth implementing before committing to any specific allocation policy, 
thereby helping avoid unnecessary optimization when feasible vaccination strategies perform similarly.
We formulate vaccine prioritization as an allocation-sensitivity problem through epidemiological necessity, $\mathcal{N}_{q}$, defined as the fixed-supply range of epidemic burdens generated by feasible vaccine allocations. 
This range distinguishes the downstream question of how to optimize prioritization from the upstream question of whether prioritization is worth optimizing. 
For policymakers, high epidemiological necessity indicates that allocation choices can materially change epidemic burden. 
In such settings, detailed modeling, tighter targeting, and transparent justification of prioritization rules are epidemiologically justified. 
Low epidemiological necessity means that feasible allocations lie within a narrow outcome band under the same vaccine supply. 
In these regimes, broad eligibility categories, simple heuristics, or population-proportional rules may be epidemiologically adequate, 
allowing policy effort to shift toward uptake, delivery speed, communication, and public trust.

Across prevention objectives, our results show that the necessity of optimization is governed by the competition between indirect protection through high-contact groups and direct protection of high-benefit groups. 
As transmission intensity increases, optimal vaccine prioritization shifts from indirect protection toward direct protection. 
When these two protection routes are balanced, marginal benefits are distributed more evenly and feasible allocations perform similarly. 
When one route dominates, marginal benefits concentrate in specific groups and allocation choices become consequential.
Age-specific objective weights, illustrated by death minimization through heterogeneous fatality risks, reshape this protection landscape and shift where optimization is most necessary. 
Optimal vaccine prioritization with different objectives shifts at different transmission intensities, 
creating a cross-objective trade-off interval where optimizing one objective substantially sacrifices another,
thereby revealing when the choice of prevention target matters most.

The death-minimization results also show how the framework extends beyond cumulative infections and deaths. 
Cumulative infection minimization assigns equal weight to infections across age groups, whereas death minimization weights infections by age-specific fatality risk. 
Many prevention targets can be represented, at least approximately, as weighted burden objectives, including hospitalization, severe disease, life-years lost, or priority-weighted infections. 
Changing the objective alters which age groups carry the largest marginal contribution to the burden being minimized. 
As a result, objective weights do not merely rescale epidemic burden; they reshape the balance between indirect and direct protection (Supplementary Section~7).
Optimization necessity therefore depends not only on how heterogeneous the weights are, but also on how high-weight groups align with the contact structure of the population. 
This makes epidemiological necessity applicable to a broader class of prevention targets while preserving the same interpretation.

The vaccine-response delay can also be interpreted through the dimensionless ratio $\delta/T_g$, which measures the number of transmission generations that elapse before vaccine-induced protection develops. 
Under the present parameterization, epidemiological necessity declines markedly when $\delta/T_g \gtrsim 2$. 
For the Pfizer--BioNTech BNT162b2 mRNA COVID-19 vaccine, protection began to emerge approximately 12 days after the first dose~\cite{Polack2020BNT162b2}, corresponding to $\delta/T_g \approx 2.4$ under our COVID-19 calibration of $T_g=5.05$ days. 
This suggests that fine-grained optimization may have limited within-wave necessity for an already expanding wave. 
However, real COVID-19 epidemics unfolded through multiple waves, so vaccination during an earlier wave or between waves can affect later waves through persistent immune protection. 
If protection generated by earlier vaccination has matured before the next wave begins, it could be represented in the initial protected state of that wave, with an effective $\delta$ close to $0$. 
If pre-existing protection has waned and booster vaccination is administered during the subsequent wave, then $\delta$ can instead represent the onset time of the incremental protection induced by the booster. 
Clinical data for booster doses of BNT162b2 show that additional protection can appear within the first week after boosting~\cite{Moreira2022ThirdDose}; 
under the same $T_g\approx 5.05$ day scale, this corresponds to $\delta/T_g \lesssim 1.4$, a range in which the necessity landscape still retains clear regime structure.
Thus, the same allocation strategy may have low within-wave necessity for an ongoing wave while retaining substantial allocation sensitivity over a longer multi-wave horizon. 
More generally, $\delta/T_g$ depends jointly on the speed of vaccine-induced protection and the pathogen transmission time scale. 
For pathogens with longer generation or serial intervals, the same calendar response delay corresponds to a smaller relative delay: 
reported intervals are approximately 18 days for rubella, mumps and smallpox, 14 days for varicella, and 11.7 days for measles~\cite{Vink2014SerialIntervals}. 
Such settings leave a wider window in which optimization of vaccine allocation can still materially affect the burden of the ongoing wave.

We use the one-time vaccination campaign as the main framework because it provides a cleaner structural diagnostic of allocation sensitivity, 
while also offering a useful limiting case for allocation-sensitive immunity carried into a later wave when earlier vaccination has already matured.
In this setup, the total vaccine supply, vaccination time, and epidemic state at vaccination are fixed; the only vaccination-control variable that changes across feasible strategies is the age distribution of the same supply. 
In the one-time vaccination campaign, epidemiological necessity measures allocation sensitivity itself. 
This simplification also makes the theoretical analysis more tractable: 
the allocation-to-burden map, marginal-benefit calculations, indirect--direct protection balance, 
and competing minima in allocation space can therefore be studied more transparently within the same fixed-supply allocation space, 
rather than being entangled with changes in epidemic state across delivery rounds.
A time-course rollout would introduce an additional delivery-tempo dimension: once doses are delivered over time, 
outcomes depend not only on the age allocation but also on the spacing between delivery rounds, the total rollout duration, 
the rollout start time, the speed of epidemic progression, and whether allocations are re-optimized as epidemic states change. 
These factors can mix the effect of allocation with the effect of rollout timing. 
We therefore use the one-time campaign to establish the main mechanisms, including the balance between indirect and direct protection, the low-necessity valley, the reshaping of death-minimizing necessity by fatality-risk heterogeneity, and the COTI. 
The time-course analysis in Supplementary Section~5 then serves as a robustness check, showing that these qualitative conclusions persist when allocation is optimized and evaluated under time-course vaccine delivery.

Several scope limitations guide the interpretation of the framework. 
First, the model stratifies the population by age, whereas real prioritization may also depend on occupation, comorbidities, geography, socioeconomic vulnerability, healthcare access, and behavioral heterogeneity. 
Extending epidemiological necessity to richer population stratifications would allow allocation sensitivity to be assessed across the additional dimensions used in real vaccination programs.
Second, epidemiological necessity is estimated conditionally on specified model inputs. 
Parameters such as transmission intensity, generation time, contact structure, vaccine supply, vaccine efficacy, response delay, and age-specific fatality risk are uncertain and may change during an epidemic. 
This uncertainty is especially relevant near transition regimes, where small parameter changes can shift the inferred balance between indirect and direct protection or alter the width of the cross-objective trade-off interval. 
Future work should propagate parameter uncertainty into $\mathcal{N}_q$ and the cross-objective penalties, allowing necessity estimates to be reported as uncertainty ranges or robust regime classifications rather than point estimates.

\section*{Methods}

\subsection*{Allocation-to-burden framework}

All objective values are evaluated through an allocation-to-burden map that takes an age-specific vaccine allocation $\boldsymbol{\vartheta}$ as input and returns the corresponding final epidemic burden. 
The notation and dynamical model used to compute this map are described in Supplementary Sections~1--2. 
For a given allocation $\boldsymbol{\vartheta}$, let $\tilde{\boldsymbol{c}}(\boldsymbol{\vartheta})$ denote the final cumulative infection vector. 
For a prevention objective $q$ with age-specific weight vector $\boldsymbol{\varpi}_q$, the corresponding final burden is:
\begin{align}
\tilde{\chi}_{q}(\boldsymbol{\vartheta}) = (\boldsymbol{\rho}\odot\boldsymbol{\varpi}_q)^{\top}\tilde{\boldsymbol{c}}(\boldsymbol{\vartheta}). \label{eq:objective_burden}
\end{align}
Cumulative infections and deaths are the two objectives considered in the main analysis, with $\boldsymbol{\varpi}_{\mathrm{c}}=\boldsymbol{1}$ and $\boldsymbol{\varpi}_{\mathrm{d}}=\boldsymbol{\epsilon}$, respectively:
\begin{align}
\tilde{\chi}_{\mathrm{c}}(\boldsymbol{\vartheta}) &= \boldsymbol{\rho}^{\top}\tilde{\boldsymbol{c}}(\boldsymbol{\vartheta}), \label{eq:cinf_burden}\\
\tilde{\chi}_{\mathrm{d}}(\boldsymbol{\vartheta}) &= (\boldsymbol{\rho}\odot\boldsymbol{\epsilon})^{\top}\tilde{\boldsymbol{c}}(\boldsymbol{\vartheta}), \label{eq:death_burden}
\end{align}
where $\boldsymbol{\rho}$ is the population distribution and $\boldsymbol{\epsilon}$ is the vector of age-specific infection fatality rates. 
Unless IFR heterogeneity is explicitly varied, $\boldsymbol{\epsilon}$ denotes the baseline age-specific IFR profile; in IFR-heterogeneity analyses, it is replaced by the interpolated profile $\boldsymbol{\epsilon}(\gamma)$ defined below.

\subsection*{Feasible allocation set and numerical optimization}

For each parameter setting and prevention objective $q$, 
the bounding allocations in Eqs.~\eqref{eq:opt_allocation}--\eqref{eq:max_allocation} are optimized over a fixed-supply feasible set. 
Let $B$ denote the fixed vaccine supply as a fraction of the total population. 
At decision time $t$, the feasible allocation set is:
\begin{align}
\Theta(B,t)=\left\{\boldsymbol{\vartheta}\in\mathbb{R}_{+}^{n}:0\le \vartheta_l\le s_l(t),\;\boldsymbol{\rho}^{\top}\boldsymbol{\vartheta}=B\right\}. \label{eq:feasible_allocation_set}
\end{align}
All allocations in $\Theta(B,t)$ use the same total vaccine supply and differ only in how doses are distributed across age groups. 
When the vaccine supply and decision time are fixed, we write $\Theta(B,t)$ simply as $\Theta$ in the Results. 
We use sequential least-squares programming (SLSQP) for the constrained optimization~\cite{2020SciPy-NMeth}. 
For minimization, the objective function is $\tilde{\chi}_q(\boldsymbol{\vartheta})$; for maximization, we minimize $-\tilde{\chi}_q(\boldsymbol{\vartheta})$ over the same feasible set. 
To reduce sensitivity to local optima, each optimization is run from multiple feasible initializations, and the best feasible solution is retained. 

\subsection*{Vaccine-response delay and epidemic tempo}

 The calendar value of $\delta$ should be interpreted within the epidemic clock of our model, rather than as a universal biological time-to-protection across outbreak time scales.
Although $R_0$ is a key determinant of epidemic scale, it does not fix the calendar speed of transmission; epidemics with the same $R_0$ can unfold on different time scales.
If the epidemic time scale is rescaled by a factor $h$ while $R_0$ is held fixed, the epidemic trajectory is preserved up to a reparameterization of time.
Preserving the same pre-protection infection risk then requires the vaccine-response delay to be rescaled in the same way, $\delta\mapsto h\delta$~\cite{andreasen2011final,kermack1927contribution,breda2012formulation,wallinga2007generation,park2019practical}.
The effect of $\delta$ on epidemiological necessity therefore reflects response timing relative to epidemic tempo~\cite{dietz1993estimation,wallinga2007generation,delamater2019complexity,park2019practical,gostic2020practical,park2021forward}.
For example, for a real epidemic that unfolds over several years, a response delay lasting several weeks may correspond to only a few days on the time scale of our model with the same $R_0$.
Thus, in the main analysis, numerical values of $\delta$ should be read as shorter or longer response-timing scenarios under the model time scale, not as universal biological calendar times to protection.

This relative timing can be summarized by the dimensionless delay $\delta^*=\delta/T_g$, where $T_g$ is the mean generation time, 
$T_g=\int_0^\infty \tau\widehat{\psi}_{\mathrm{gen}}(\tau)\,d\tau$
and $\widehat{\psi}_{\mathrm{gen}}$ is the normalized generation-time distribution.
Under the COVID-19 calibration, $\widehat{\psi}_{\mathrm{gen}}$ is a Weibull distribution with shape $\alpha_{\mathrm{gen}}=2.826$ and scale $\beta_{\mathrm{gen}}=5.665$~\cite{ferretti2020quantifying}.
For a Weibull generation-time density, the mean-generation-time integral above evaluates to $T_g=\beta_{\mathrm{gen}}\Gamma(1+1/\alpha_{\mathrm{gen}})$, giving $T_g=5.05$ days under these parameter values.

\subsection*{Protection orientation}

We quantify whether an allocation is oriented toward indirect or direct protection using a signed protection-orientation index. 
The indirect-protection reference direction is $\boldsymbol{A}\boldsymbol{\rho}$, which represents contact opportunities. 
The direct-protection reference direction is $\boldsymbol{\varsigma}_q\odot\boldsymbol{\rho}$, where $\boldsymbol{\varsigma}_q$ encodes the objective-specific direct benefit of vaccination. 
At the fixed decision time, we write $\boldsymbol{\sigma}\equiv\boldsymbol{\sigma}(t)$ for the probability of avoiding infection during the vaccine-response window. 
For cumulative infections, $\boldsymbol{\varsigma}_{\mathrm{c}}=\boldsymbol{\sigma}$; for deaths, $\boldsymbol{\varsigma}_{\mathrm{d}}=\boldsymbol{\sigma}\odot\boldsymbol{\epsilon}$.

For an allocation $\boldsymbol{\vartheta}$, we define:
\begin{align}
\xi_q(\boldsymbol{\vartheta}) = \frac{\varphi(\boldsymbol{\vartheta},\boldsymbol{A}\boldsymbol{\rho})-\varphi(\boldsymbol{\vartheta},\boldsymbol{\varsigma}_q\odot\boldsymbol{\rho})}{\varphi(\boldsymbol{A}\boldsymbol{\rho},\boldsymbol{\varsigma}_q\odot\boldsymbol{\rho})}, \label{eq:protection_orientation}
\end{align}
where
\begin{align}
\varphi(\boldsymbol{x},\boldsymbol{y}) = \arccos\left(\frac{\boldsymbol{x}^{\top}\boldsymbol{y}}{\|\boldsymbol{x}\|_2\|\boldsymbol{y}\|_2}\right). \label{eq:angular_distance}
\end{align}
With this convention, $\xi_q<0$ indicates an indirect-protection orientation, $\xi_q>0$ indicates a direct-protection orientation, and $\xi_q\approx0$ indicates balanced protection.
For objective-specific bounding allocations, we use the shorthand $\xi^{\mathrm{opt}}_q=\xi_q(\boldsymbol{\vartheta}_{q}^{\mathrm{opt}})$ and $\xi^{\mathrm{max}}_q=\xi_q(\boldsymbol{\vartheta}_{q}^{\mathrm{max}})$, where the subscript denotes the prevention objective and the superscript denotes the allocation type.

\subsection*{Marginal vaccination benefits}
We define the marginal vaccination benefit of an age group as the reduction in final burden produced by assigning an infinitesimal dose to that group. 
Here we retain the time subscript in $\boldsymbol{\vartheta}_t$ to emphasize that the derivative is taken with respect to the allocation made at decision time $t$.
Let $\boldsymbol{J}_t=\partial\tilde{\boldsymbol{c}}/\partial\boldsymbol{\vartheta}_t$ be the Jacobian of final cumulative infections with respect to allocation at time $t$. 
For an objective with age-specific weight vector $\boldsymbol{\varpi}_q$, where $\boldsymbol{\varpi}_{\mathrm{c}}=\boldsymbol{1}$ for cumulative infections and $\boldsymbol{\varpi}_{\mathrm{d}}=\boldsymbol{\epsilon}$ for deaths, the marginal benefit vector is:
\begin{align}
\boldsymbol{\mu}^{q}_{t} = -\left.\boldsymbol{J}_t^{\top}(\boldsymbol{\rho}\odot\boldsymbol{\varpi}_q)\oslash\boldsymbol{\rho}\right|_{\boldsymbol{\vartheta}_t=\boldsymbol{0}}. \label{eq:marginal_benefit}
\end{align}
We summarize age-group heterogeneity in marginal vaccination benefit using the Gini coefficient. 
For the nonnegative marginal-benefit vector $\boldsymbol{\mu}^{q}_{t}$, the Gini coefficient is:
\begin{align}
	G(\boldsymbol{\mu}^{q}_{t}) = \frac{\sum_{l=1}^{n}\sum_{m=1}^{n}|\mu^{q}_{t,l}-\mu^{q}_{t,m}|}{2n\sum_{l=1}^{n}\mu^{q}_{t,l}}. \label{eq:marginal_benefit_gini}
\end{align}
When all entries are zero, we define $G(\boldsymbol{\mu}^{q}_{t})=0$, corresponding to no distinguishable heterogeneity across age groups. 

\subsection*{IFR heterogeneity}

To vary IFR heterogeneity, we interpolate between a homogeneous IFR profile and the baseline age-specific IFR profile $\widehat{\boldsymbol{\epsilon}}$. 
Let $\bar{\boldsymbol{\epsilon}}$ be the homogeneous profile with all entries equal to the mean IFR. 
For $\gamma\in[0,1]$, we define:
\begin{align}
\boldsymbol{\epsilon}(\gamma) = \bar{\boldsymbol{\epsilon}}+\gamma\left(\widehat{\boldsymbol{\epsilon}}-\bar{\boldsymbol{\epsilon}}\right). \label{eq:ifr_heterogeneity}
\end{align}
Thus, $\gamma=0$ gives a homogeneous IFR profile, and $\gamma=1$ gives the baseline age-specific IFR profile. 
This interpolation preserves the mean IFR while varying how strongly fatality risk differs across age groups. 
For death minimization, the resulting profile is used as the objective weight, $\boldsymbol{\varpi}_{\mathrm{d}}=\boldsymbol{\epsilon}(\gamma)$, whereas cumulative-infection minimization uses the uniform weight $\boldsymbol{\varpi}_{\mathrm{c}}=\boldsymbol{1}$.

\subsection*{Cross-objective allocation correlation}

The cross-objective analysis compares the infection-minimizing allocation $\boldsymbol{\vartheta}^{\mathrm{opt}}_{\mathrm{c}}$ with the death-minimizing allocation $\boldsymbol{\vartheta}^{\mathrm{opt}}_{\mathrm{d}}$. 
We quantify their cross-objective allocation correlation using the Pearson correlation coefficient between their age-specific allocation profiles:
\begin{align}
	\theta(\boldsymbol{\vartheta}^{\mathrm{opt}}_{\mathrm{c}},\boldsymbol{\vartheta}^{\mathrm{opt}}_{\mathrm{d}}) = \frac{\sum_{l=1}^{n}(\vartheta_{\mathrm{c},l}^{\mathrm{opt}}-\bar{\vartheta}_{\mathrm{c}}^{\mathrm{opt}})(\vartheta_{\mathrm{d},l}^{\mathrm{opt}}-\bar{\vartheta}_{\mathrm{d}}^{\mathrm{opt}})}{\sqrt{\sum_{l=1}^{n}(\vartheta_{\mathrm{c},l}^{\mathrm{opt}}-\bar{\vartheta}_{\mathrm{c}}^{\mathrm{opt}})^2}\sqrt{\sum_{l=1}^{n}(\vartheta_{\mathrm{d},l}^{\mathrm{opt}}-\bar{\vartheta}_{\mathrm{d}}^{\mathrm{opt}})^2}}. \label{eq:allocation_correlation}
\end{align}
Here $\bar{\vartheta}_{q}^{\mathrm{opt}}=n^{-1}\sum_{l=1}^{n}\vartheta_{q,l}^{\mathrm{opt}}$. 
Low $\theta$ indicates that the infection- and death-minimizing allocations no longer prioritize age groups in a similar pattern.

\section*{Data Availability}

\vspace*{-0.1in}
All relevant data are available at \url{https://github.com/fengmi9312/Necessity-of-Optimizing-Vaccine-Prioritization/releases/tag/ResultData}.

\vspace*{-0.1in}
\section*{Code Availability}

\vspace*{-0.1in}
The GitHub repository, which includes the source code for all the figure results, can be accessed at
\url{https://github.com/fengmi9312/Necessity-of-Optimizing-Vaccine-Prioritization}.

\vspace*{-0.1in}
\section*{Acknowledgments}
\vspace*{-0.1in}
M.F. acknowledges support from the National Natural Science Foundation of China (NSFC) through the Young Scientists Fund (Category C, Grant No.~12505050). C.-S.Z. acknowledges support from the Hong Kong Research Grants Council (Grant Nos.~SRFS2324-2S05, GRF12202124, and C4012-22G), the Guangdong and Hong Kong Universities ``1+1+1'' Joint Research Collaboration Scheme (Grant No.~2025A0505000011), and the Hong Kong Baptist University Research Committee (Grant No.~RC-SFCRG/23-24/SCI/06). L.T. acknowledges support from the Hong Kong Research Grants Council (Grant Nos.~C2005-22Y, 12301723, and 12301624), the Hong Kong Chinese Medicine Development Fund (Grant No.~22B2/049A), the National Natural Science Foundation of China (Grant No.~12275229), the Guangdong and Hong Kong Universities ``1+1+1'' Joint Research Collaboration Scheme, Hong Kong Baptist University's Chinese Medicine Research in Conjunction with the Chinese Medicine Hospital (Grant No.~TFD2025-P05), and the Hong Kong Baptist University Research Committee (Grant Nos.~RC-FNRA-IG/23-24/SCI/05 and CRMS/23-24/03). We thank Prof.~Matthew Turner (University of Warwick) for helpful comments on the manuscript.

\vspace*{-0.1in}
\section*{Author Contributions}
\vspace*{-0.1in}
M.F., L.T. and C.-S.Z. designed research; M.F. performed research; L.T. and C.-S.Z. contributed analytic tools; M.F., L.T. and C.-S.Z. analyzed data; M.F., Z.-H.L., L.T. and C.-S.Z. discussed the results and wrote the paper.

\vspace*{-0.1in}
\section*{Competing Interests}

\vspace*{-0.1in}
The authors declare no competing interests.

\vspace*{-0.1in}
\section*{Correspondence}

\vspace*{-0.1in}
To whom correspondence should be addressed: cszhou@hkbu.edu.hk, liangtian@hkbu.edu.hk

\end{document}